\documentclass[prd,aps,letterpaper,floatfix,superscriptaddress,preprintnumbers,twocolumn,10pt,nofootinbib]{revtex4-1}
\pdfoutput=1

\usepackage{dcolumn}
\usepackage{bm}
\usepackage[dvips]{graphicx}
\usepackage{epsfig}
\usepackage{amsmath, amsfonts, amssymb, slashed}
\usepackage{slashed}
\usepackage{xcolor}
\usepackage{bm}        
\usepackage{graphicx,psfrag,subfigure}
\usepackage{booktabs}
\usepackage[colorlinks]{hyperref}
\usepackage{units}
\usepackage{float}

\usepackage{ulem}
\hypersetup{
    colorlinks=true,
    linkcolor=blue,
    filecolor=magenta,      
    urlcolor=cyan,
}
\def\lsim{\mathrel{\raise.3ex\hbox{$<$\kern-.75em\lower1ex\hbox{$\sim$}}}}
\def\gsim{\mathrel{\raise.3ex\hbox{$>$\kern-.75em\lower1ex\hbox{$\sim$}}}}

\definecolor{red}{rgb}{1.0, 0, 0}

\newcommand{\beqa}{\begin{eqnarray}}
\newcommand{\eeqa}{\end{eqnarray}}

\newcommand{\be}{\begin{equation}}
\newcommand{\ee}{\end{equation}}

\graphicspath{{Images/}}

\begin{document}
\title{Convolutional Neural Networks for Direct Detection of Dark Matter}
\author{Charanjit K. Khosa} 
\email{charanjit.kaur@sussex.ac.uk}
\affiliation{Department of Physics and Astronomy, University of Sussex, 
Brighton BN1 9QH, UK}
\author{Lucy Mars}
\email{lucy.c.mars1@gmail.com}
\affiliation{Department of Physics and Astronomy, University of Sussex, 
Brighton BN1 9QH, UK}
\author{Joel Richards}
\email{jacr20@sussex.ac.uk}
\affiliation{Department of Physics and Astronomy, University of Sussex, 
Brighton BN1 9QH, UK}
\author{Veronica Sanz} 
\email{V.Sanz@sussex.ac.uk}
\affiliation{Department of Physics and Astronomy, University of Sussex, 
Brighton BN1 9QH, UK}
\affiliation{Alan Turing Institute, British Library, 96 Euston Road, London NW1 2DB, UK}
\affiliation{Instituto de F\'isica Corpuscular (IFIC), Universidad de Valencia-CSIC, E-46980, Valencia, Spain}

\date{\today}
\begin{abstract}
The XENON1T experiment uses a time projection chamber (TPC) with liquid Xenon to search for Weakly Interacting Massive Particles (WIMPs), a proposed Dark Matter particle, via direct detection. As this experiment relies on capturing rare events, the focus is on achieving a high recall of WIMP events. Hence the ability to distinguish between WIMP and the background is extremely important. To accomplish this, we suggest using Convolutional Neural Networks (CNNs); a Machine Learning procedure mainly used in image recognition tasks. To explore this technique we use XENON collaboration open-source software to simulate the TPC graphical output of Dark Matter signals and main backgrounds.  A CNN turns out to be a suitable tool for this purpose, as it can identify features in the images that differentiate the two types of events without the need to manipulate or remove data in order to focus on a particular region of  the detector.  We find that the CNN can distinguish between the dominant background events (ER) and 500 GeV WIMP events with a recall of 93.4\%, precision of 81.2\% and an accuracy of 87.2\%. 
\end{abstract}
\maketitle 

\section{Introduction}
The mystery of Dark Matter is one of the main motivations to search for physics beyond the Standard Model of Particle Physics. The detection of Dark Matter interactions beyond gravitational would be a crucial step for both the Particle and Astroparticle Physics communities. Direct Detection experiments, such as XENON1T, search for instances where Dark Matter particles scatter with Standard Model (SM) particles and transfer energy to them inside a detector.

For particle Dark Matter direct detection experiments, the main building blocks are a cryogenic material single-phase time projection chamber (TPC), a dual-phase TPC, or bubble chambers. Among the leading direct detection experiments, a TPC is used in DarkSide-50 \cite{Agnes:2018fwg}, LUX \cite{Akerib:2016vxi}, PandaX-II \cite{Cui:2017nnn} and XENON1T~\cite{Aprile:2017aty}. In this work, we focus on the output of TPC as a part of the XENON1T experiment. The light signals recorded by the photomultiplier tubes (PMTs) due to the prompt scintillation and secondary scintillation are used to infer the type of interactions, namely to distinguish WIMP and background events. The dominant background sources are beta particles, neutrons and gamma-ray photons. 

Typically, in order to achieve a large signal-to-background ratio in the data, one requires a substantial number of cuts to the data to be performed based on certain discrimination parameters. To squeeze as much signal as possible, it is crucial to improve on recall of anomalous signal events, and more effective limits to these discrimination parameters are sought in an attempt to improve the detection probability of such anomalous events.

Machine Learning (ML) methods based analysis strategy may provide a unique and flexible alternative to profile likelihood approach often used in experimental analyses for the  signal identification. In particular, Deep Learning models such as Convolutional Neural Networks (CNN) are able to heuristically learn patterns in highly-complex input space, leading to an ability to detect anomalous signals without the need to manipulate or remove as much data. In this paper, a novel, Deep Learning model is developed using a CNN which can discriminate between simulated background and WIMP waveform and hitpattern images from the XENON1T experiment. 

In the field of Particle and Astrophysics, Deep Learning is showing promising results (see e.g. \cite{Carleo:2019ptp}). Convolution Neural Networks has also been found very efficient for simulating the Dark Matter in N-body simulations of the galaxies \cite{Zhang:2019ryt}.  They also offer improved sensitivity for cosmological observations from weak lensing maps~\cite{Fluri:2019qtp}. Machine learning shows promising reach for disentangling among collider Dark Matter searches~\cite{Khosa:2019kxd}, using substructure based Dark Matter probes for non-collider  searches~\cite{Brehmer:2019jyt,Alexander:2019puy,DiazRivero:2019hxf}, and for cosmological Dark Matter~\cite{cosmoDM}. 

This paper is organised as follows. In Section~\ref{NoEvents} we briefly describe the XENON1T experiment,  explaining how the time projection chamber is used to look for proposed WIMP events and the types of backgrounds. Section~\ref{simulations} is devoted to the process of simulating Dark Matter and electronic recoil background events using the open source data processing software created by the XENON collaboration. The architecture of the CNN used is explained in section \ref{cnn}, as well as the training and testing procedure for the model. In the last section~\ref{conclusions}, we discuss the results.

\section{WIMP and Background Events at XENON1T\label{NoEvents}}

The XENON experiment is an underground research facility, operated at the Laboratori Nazionali del Gran Sasso (LNGS) in Italy. Starting in 2006, the XENON10 experiment used a 25 kg (14 kg target mass) Xenon dual phase time projection chamber to search for WIMPs \cite{Aprile:2010bt}. This was followed by the XENON100 experiment in 2008, containing 62 kg target mass of LXe (161 kg total) \cite{Aprile:2011dd}. The most recent experiment is the XENON1T experiment, a 3.2 tonne LXeTPC with a fiducial volume of roughly 2 tonnes \cite{Aprile:2017aty}. The TPC was designed to detect nuclear recoils (NRs) from WIMP particles scattering off the Xe nuclei \cite{Aprile:2017iyp}.

\begin{figure}[htp]
    \centering
    \includegraphics[angle=0,width=1.\linewidth,trim={0mm 0mm 0mm 0mm},clip]{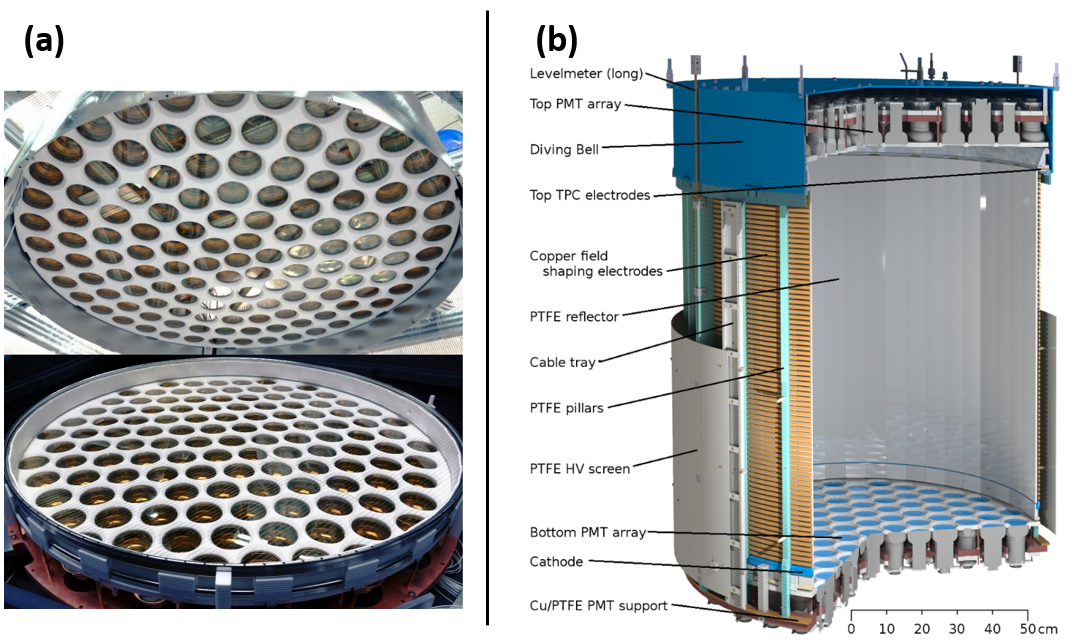}
    \caption{(a) The top (127) and bottom (121) PMT arrays used in the XENON1T TPC. (b) Illustration of the XENON1T TPC. Figure taken from~\cite{Aprile:2017aty}.}
    \label{fig:Experiment}
\end{figure}

A vital part of a TPC are the photomultiplier tubes. When a photon hits the photocathode in the PMT it produces electrons which are then accelerated by a high-voltage field. The number of electrons increases within a chain of dynodes by secondary emission. A total of 248 PMTs are used in the TPC to record signals. They are split into the top array (which contains 127 PMTs) and bottom array (121 PMTs), in order to achieve a uniform field and allow good position reconstruction. Fig. \ref{fig:Experiment} shows the PMTs and the TPC used in the XENON1T experiment. More information about the XENON1T TPC can be found in \cite{Aprile:2017aty}. Figure \ref{fig:Experiment} shows the PMTs and TPC used in XENON1T.

If an incoming particle interacted with the liquid Xenon, it would produce a scintillation of light and ionization. The S1 signal is the light seen by the top and bottom PMTs (due to total internal reflection at the boundary). The electron charges that were released during ionisation then drift upwards towards the gaseous Xenon (GXe) due to the electric field between the cathode gate and anode. They are then extracted by a stronger extraction field, E$_{extraction}$, creating another larger scintillation light signal (S2) seen by the top PMTs. The position of the original interaction can be reconstructed in 3 dimensions by using the S2 signal pattern (giving the lateral position) and the difference in time between S1 and S2 (depth of interaction)~\cite{Aprile:2017aty}.

However, there are many other processes that can lead to the creation of a light signal within the XENON1T experiment. For example, in order to reduce cosmic rays reaching the Xenon, the experiment was carried out deep underground, under the Italian Apennines. To reduce natural background radioactivity the TPC was submerged in a water tank and made from  material of low natural radioactivity. Furthermore, only events that occurred within the inner tonne of LXe were used, allowing for the rest of the LXe to be used as more shielding. This was possible due to the large mass density of LXe (almost 3000 kg/m$^3$) and its high atomic number, meaning particles, such as gammas, can only travel for a short distance before being stopped \cite{Pelssers}.

Despite the shielding, there are six types of background events that can be detected within the search region. Table \ref{table:Background} shows the expected number of each of these background events, as well as the expected number of events for a spin-dependent 500 GeV/$c^2$ WIMP with a cross-section of $10^{-45}$ cm$^2$, over the time period of the first science run of XENON1T (34.2 live-days). This WIMP benchmark is chosen among the allowed values by direct, indirect and collider Dark Matter searches~\cite{WIMPbanchmark}. The expected number of events were calculated using {\tt Laidbax}~\cite{Laidbax} and the {\tt blueice} Monte Carlo model~\cite{Blueice}. These results agree with the XENON1T collaboration paper (\cite{Aprile:2017iyp}), however, they used a WIMP mass of 50 GeV/$c^2$ and a $10^{-46}$ cm$^2$ cross-section, and hence have a different expected number of WIMP events.

The particles in the detector release energy in the form of a nuclear or electronic recoil (NR or ER). This means the incoming particle will either scatter directly from the nucleus of the target atom, or it will interact with the electron cloud \cite{Pelssers}. Since the energy transferred between a WIMP and a XENON nucleus is much higher than for any electronic recoil (due to kinematics), most direct detection dark matter experiment are built to search for WIMP NRs. Therefore, our ability to differentiate between NR and ER is very important when searching for WIMPs. Table \ref{table:Background} shows that the number of expected WIMP events is much higher than for any of the NR background.

\begin{table}[H]
	\centering 
	\begin{tabular}{c c } 
		\hline 
		Name	& Expected number of events \\
		\hline
		Electronic recoils (ER) & 61.879487 \\
		CNNS ($\nu$) & 0.000901 \\
		Radiogenic neutrons & 0.058570  \\	
		Accidental coincidences (acc) & 0.220000  \\		
		Wall leakage (wall) & 0.520000 \\
		Anomalous (anom) & 0.090004 \\	
		500 GeV/$c^2$, $10^{-45}$ cm$^2$ WIMP & 35.029005 \\
		\hline
	\end{tabular}
		\caption{Expected number of events for each type of background  over the time period of the first science run of XENON1T (34.2 live-days) within the fiducial mass and a 500 GeV/$c^2$, $10^{-45}$ cm$^2$ WIMP (Generated using {\tt Laidbax}).}
	\label{table:Background}
\end{table}

The XENON1T was designed as an ultra-low background experiment including high rejection of ER backgrounds. Even when set to 50\% NR acceptance the XENON1T could detect WIMP-like events while also rejecting roughly 99.8\% of all the ER events since it is reducible background (the same is not true for irreducible backgrounds\footnote{Other background sources are discussed in Ref. \cite{Aalbers:2018mfc}).}. The main source of the ER background is the beta decays of $^{85}$Kr and $^{214}$Pb, which causes a flat energy spectrum within the interested range \cite{Aalbers:2018mfc}.

In this work, we have focused on the classification of WIMPs with ER background using a supervised machine learning approach. The dominant ER background allows us to use balanced data sets. As this is the first step to build an alternative machine learning-based signal search for Xenon experiments, it is important to use supervised methods to understand the data behaviour and to establish benchmarks for the further steps of the analysis pipeline.

\section{Event Simulation\label{simulations}} 
In the following we introduce the three types of images that were used during our analysis,  and explain how we generated them using {\tt PaX} (Processor for Analysing XENON) software, created by the XENON collaboration \cite{PAX}~\footnote{Note that in \cite{Simola:2018ntn}, a new machine learning algorithm was used for the accurate reconstruction of 2-D interaction in the TPC.}.

First, we generate the energy spectrum of the Dark Matter particle, using {\tt wimprates}~\cite{wimprate}. Given the proposed mass and cross-section of a WIMP, {\tt wimprates} produces the differential rates of a WIMP-nucleus scattering in the standard halo model, within a liquid Xenon detector \cite{wimprate}. In this work, we used a WIMP mass of 500 $GeV^{-1}$ and a cross-section of $10^{-45} cm^{-2}$ for illustration.  More details on how to set-up the simulation are given in App.~\ref{setup}.

\begin{figure}[htp]
    \centering
    \includegraphics[angle=0,width=.9\linewidth,trim={0mm 0mm 0mm 0mm},clip]{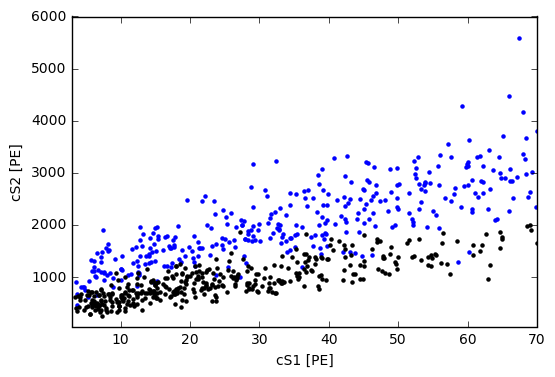}
    \caption{Illustrative plot of cS1 and cS2 for ER (blue) and WIMP (black)
    interactions.}
    \label{fig:csc}
\end{figure}

Next, {\tt Laidbax} (Likelihood- And Interpolated Density Based Analysis for XENON) \cite{Laidbax} is used to convert the energy spectrum into a model that is compatible with the {\tt PaX}~\cite{PAX} software which we use to create the graphical output of the simulated WIMP and background events. {\tt PaX} has an in-built simulator called {\tt FaX} (Fake XENON) which requires an input {\it csv} file consisting of a set of variables for each interaction to construct the waveform of the event. These variables are: instruction (simply a number given to each interaction); recoil type (NR or ER); x-position, y-position and depth of the interaction (in cm); number of photons produced; number of electrons produced; time of the interaction (in ns).

\begin{figure*}[htp]
    \centering
    \includegraphics[angle=0,width=.95\linewidth,trim={0mm 0mm 0mm 0mm},clip]{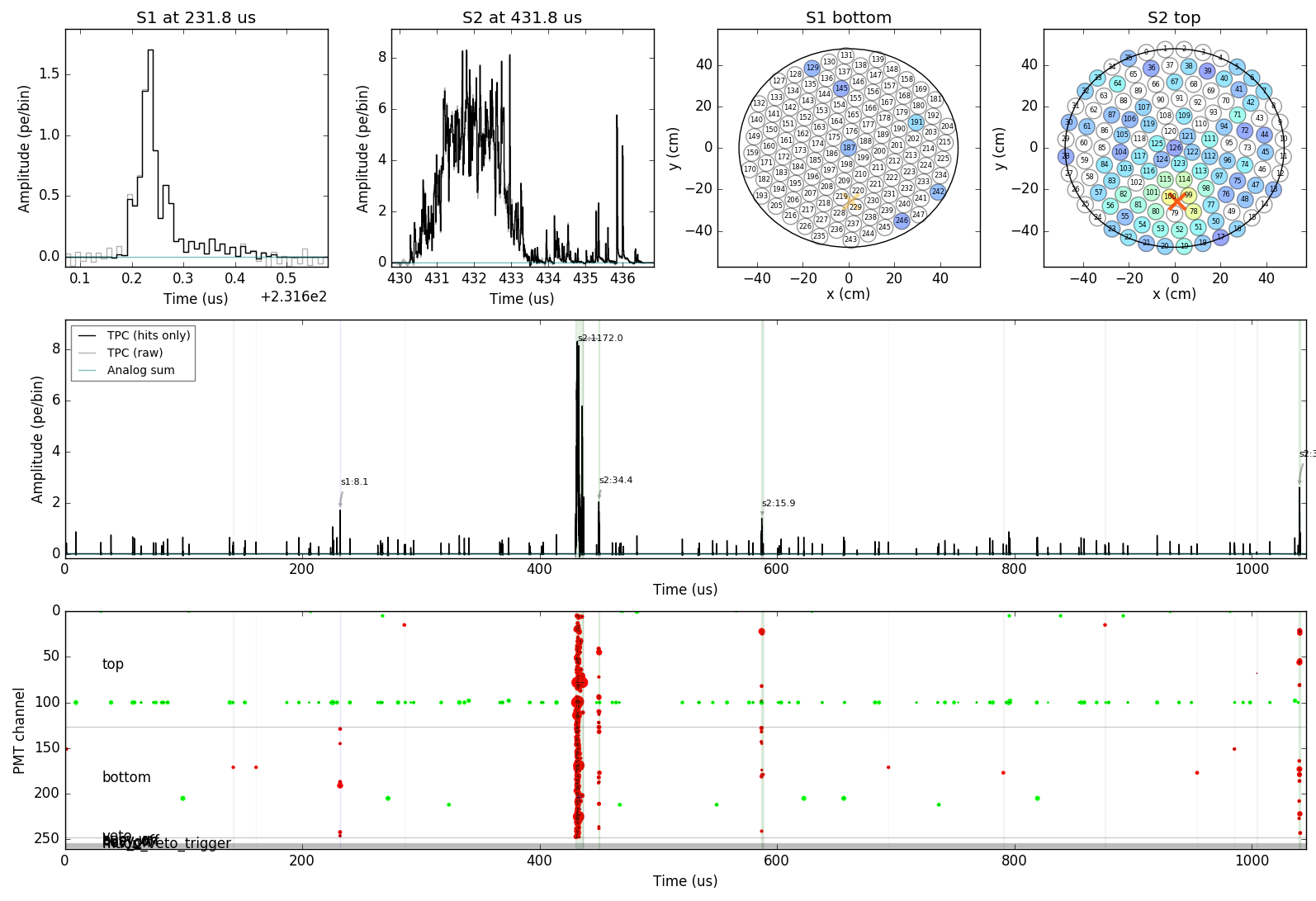}
    \caption{Example of the graphical output of {\tt PaX} for a simulated WIMP event. Top two plots on the left: largest S1 and S2 peaks. Top two plots on the right: hitpatterns for the top and bottom PMTs. Middle plot: S1 and S2 peaks in the event. Bottom plot: Red dots represent PMT channels that have detected coincident signals, whilst the green dots represent signals detected in a lone PMT.}
    \label{fig:Event0}
\end{figure*}

The ER background energy spectrum we used came pre-built in {\tt Laidbax}. Note, though, that the models used in {\tt Laidbax} are not the official models approved by the XENON collaboration, which include several more uncertainties. For example, the ER background model is a polynomial ER fit~\cite{Aalbers:2018mfc} and the yields for the NR background model are specified by the parameterisation of the global fit found in \cite{Lenardo:2014cva}.  The {\tt Laidbax} model produces a database of simulated interactions. The parameters of the interactions include; radial distance $r^2$, angle $\theta$, z-coordinate, number of photons produced, and number of electrons produced. 

When reconstructing the energy deposits of particles interacting with the LXe, the S1 and S2 signals need to be corrected in case of any time or spatial dependent signal losses. The light signal, S1, is corrected for the (x,y,z)-dependent light collection efficiency in the TPC (cS1) \cite{Breur}. The charge signal, S2, is corrected for the time and depth dependent electron lifetime since electrons can be lost while drifting in the LXe if they attach to impurities within the Xenon (cS2)~\cite{Breur}. The formulas for the corrected values can be found in Ref. \cite{Aprile:2019dme}. For illustrative purposes in Fig.~\ref{fig:csc} we plot the distribution of these events for ER and WIMP interactions. We can see the overlap between these two types of events.  

Later on in the analysis, we use raw S1 and S2 signals for two reasons. First to disentangle the dependence of the classification accuracy on the correction factors, and second because the more realistic correction maps are being implemented in the other Xenon collaboration packages, {\tt LAX} (Lichens for Analyzing XENON) and {\tt HAX} (Handy Analysis for XENON). The electron lifetime was also artificially increased to a value of $100,000 us$, further reducing the need to correct the signals. In future studies, however, the input data could be corrected using those new packages.

To convert the {\tt Laidbax} model to the input file for FaX, a series of calculations and assumptions have to be made. For example, the polar coordinates need to be converted to cartesian coordinates. The time variable, the time of the interaction in nanoseconds since the start of the event, is not given by {\tt Laidbax}. It is therefore assumed that the actual time of the interaction is not relevant, as long as it occurred within a particular time range that an event is defined by. 

\newpage
\subsection{Image Processing}
\begin{figure}[htbp!]
    \centering
    \includegraphics[angle=0,width=.8\linewidth,trim={0mm 0mm 0mm 0mm},clip]{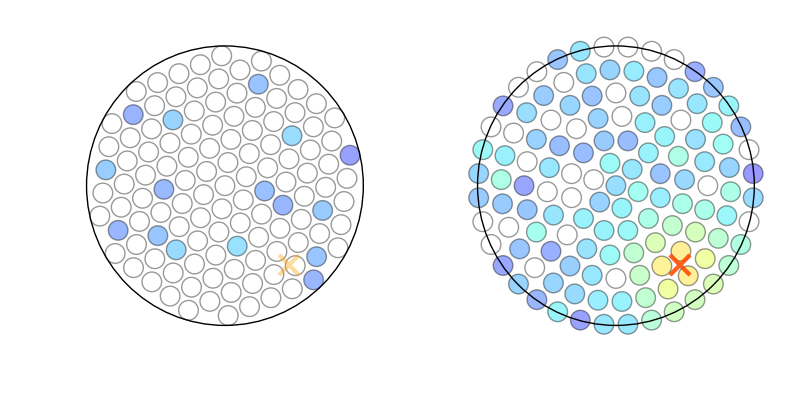}
    \caption{Example of the edited 800$\times$400 hitpattern image from a simulated WIMP event. The axis labels and numbers are removed so all images are in the same style when used in the CNN.}
    \label{fig:Hit}
\end{figure}

\begin{figure}[htbp!]
    \centering
    \includegraphics[angle=0,width=.8\linewidth,trim={0mm 0mm 0mm 0mm},clip]{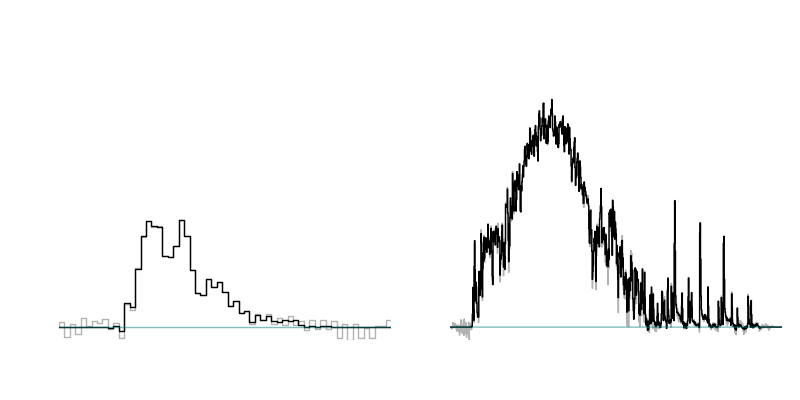}
    \caption{Example of the edited 800$\times$400 peak image from a simulated WIMP event. The axis labels and numbers are removed so all images are in the same style when used in the CNN. The y-axis are both log-scaled.}
    \label{fig:Peak}
\end{figure}

\begin{figure}[htp]
    \centering
    \includegraphics[angle=0,width=.8\linewidth,trim={0mm 0mm 0mm 0mm},clip]{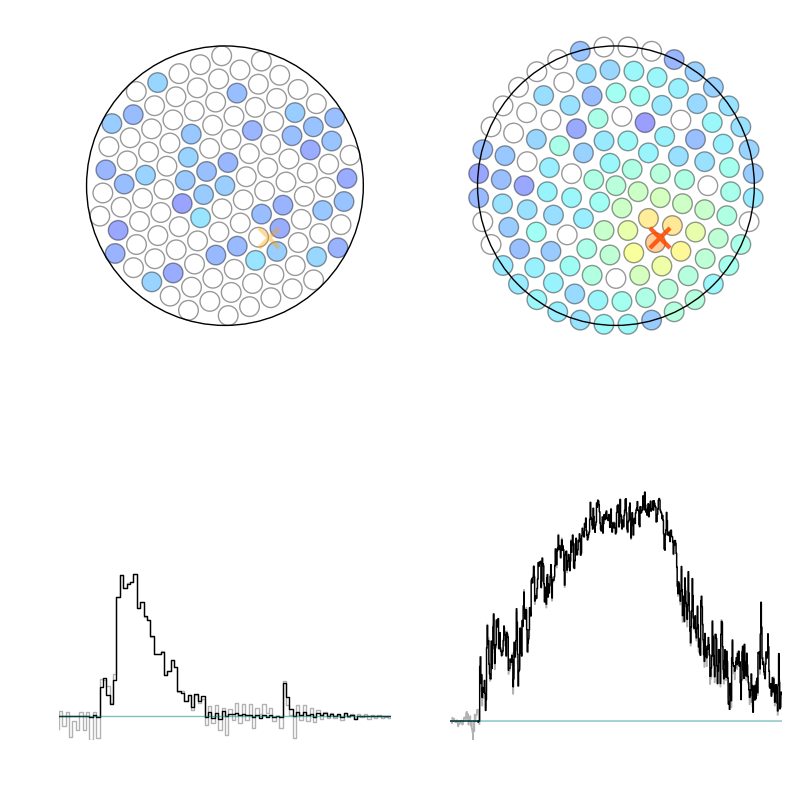}
    \caption{Example of a 800$\times$800 HitPeak image showing both the hitpatterns and largest peaks together from a simulated WIMP event. The axis labels and numbers are removed so all images are in the same format when used in the CNN.}
    \label{fig:WIMP}
\end{figure}

\begin{figure}[htp]
    \centering
    \includegraphics[angle=0,width=.8\linewidth,trim={0mm 0mm 0mm 0mm},clip]{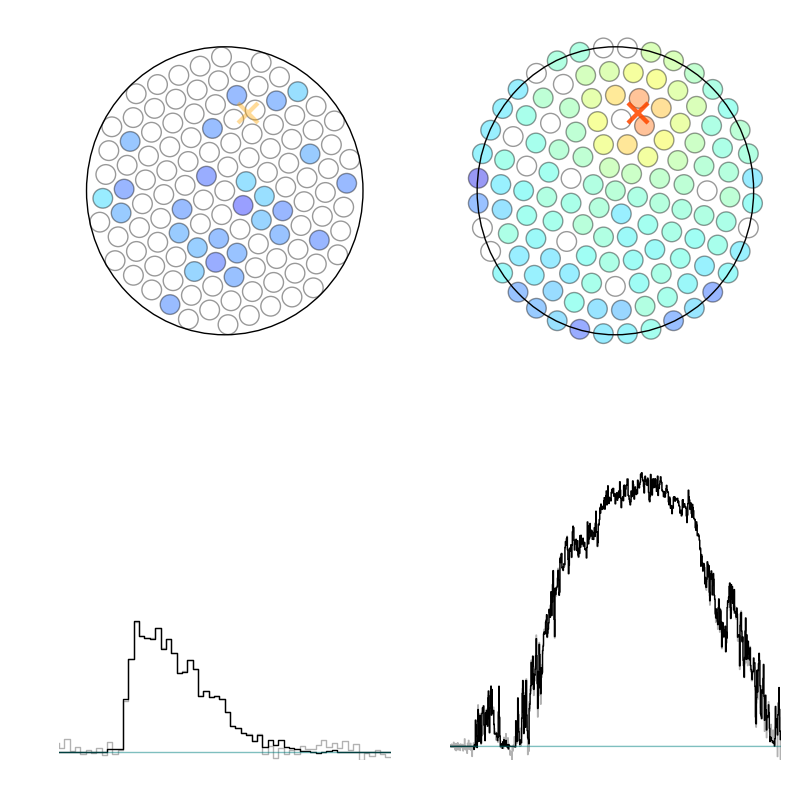}
    \caption{Example of a 800$\times$800 HitPeak image for an ER event.}
    \label{fig:ER}
\end{figure}

The original graphical output of {\tt PaX}~\cite{PAX} is shown in Fig. \ref{fig:Event0}. The images show the largest S1 and S2 peaks (top left), the hitpatterns for the top and bottom PMTs (top right), all the S1 and S2 peaks in the event (middle), and the PMTs that detected the signal (bottom). 
In order to use the images as an input to the CNN, we edited them to remove unnecessary features that are the same in every image so that the CNN can focus on features that are unique to a WIMP or background event. First, we looked at whether we could differentiate a WIMP event from the background just using the S1 and S2 hitpatterns. The `Hit' images were edited to remove the text labels in order to increase effective learning. The final image produced is 800$\times$400 pixels and an example is shown in Fig. \ref{fig:Hit}. Next, we looked at the largest S1 and S2 peak graphs.  Fig. \ref{fig:Peak} shows an example of the edited peaks for a WIMP event. The text and axis were again removed from the `Peak' plots and the y-axis was set to a log-scale with a range of 0 to 10$^{2}$. This was to show the relative pulse height between the S1 and S2 peaks since the S1/S2 size ratio is an important factor for discrimination between NR and ER \cite{Aprile:2018dbl}. Finally, we combined the hitpattern and peak graphs into one `HitPeak' image of size 800$\times$800 (shown in Fig. \ref{fig:WIMP}).  A similar image for an ER event is shown in Fig. \ref{fig:ER}; comparing this to Fig \ref{fig:WIMP} we can see that both events are visually very similar, hence the need for Machine Learning to differentiate them.

\section{CNN Architecture\label{cnn}}
This section discusses the details of  a Convolutional Neural Network used in this work (python code can be found in the github repository \cite{lucygithub}).

The majority of Machine Learning problems involve a dataset \textit{\textbf{X}=$\{$($y_{i}$,$x_{i}$), i=1,...,N$\}$}, a model \textit{g(\textbf{$\theta$})} with parameters \textit{\textbf{$\theta$}}, and a cost function \textit{C(\textbf{X},g(\textbf{$\theta$}))} (also known as the loss function). The cost function allows you to judge how well the model performs on the given dataset. A model is fitted by calculating the value of \textit{\textbf{$\theta$}} that minimises the cost function.

The most common way to minimise the cost function is to use \textit{Gradient Descent}; an algorithm that finds the local or global minima of a function. The parameters are adjusted in the direction where the gradient of the cost function is large and negative, and then the gradient is recalculated in the new position. After each iteration the model gradually converges towards a minimum (where any changes to the parameters will produce little or no change in the loss) resulting in the weights being optimised. Given the cost function \textit{C($\theta_{i}$)}, it simultaneously updates for each \textit{i = 0,...,n} until convergence is reached: 
\begin{equation}
\theta_i := \theta_i - \eta \nabla_{\theta} C(\theta_0,...,\theta_n)
\label{eq:Gradient}
\end{equation}
The learning rate, $\eta$, controls how large each step is taken during gradient descent.

Neural networks (NN) contain multiple neurons\footnote{A computational unit which performs a nonlinear transformation of its input.} stacked into hidden layers. The output of each layer then serves as the input for the following one. Each neuron takes a vector of inputs, \textbf{x}, and produces a scalar output \textit{a$_i$}(\textbf{x}). The function \textit{a$_i$} depends on the NN but it can be separated into a linear operation (which weighs the importance of the inputs) and a non-linear transformation (performed by an activation function). 

A NN calculates the gradient of the cost function using backpropagation. This algorithm contains a forward pass (going from the input layer to output layer), calculates the weighted inputs and activations for all the neurons, and then backpropagates the error (output to input layer), calculating the gradients.

A Convolutional Neural Network is a type of Neural Network Machine Learning algorithm that primarily takes images as its input and assigns weights and biases to different parts of the image. A CNN is comprised of many layers of different types, including \textit{Convolutional Layers}, \textit{Pooling Layers}, and \textit{Fully-Connected (FC) Layers}. The convolutional layer is used to extract features from the input image by passing a filter (kernel) over the image matrix. To perform different operations on the image, such as edge detection or sharpening, different types of filters are used. The layer outputs the image matrix at a reduced volume, depending on the size of the filter. A nonlinear activation function is applied after each Convolutional layer. In this paper we used a Leaky ReLU (Rectified Linear Unit) activation function. Mathematically Leaky ReLU is defined as:
\[ 
    f(x) =
    \begin{cases} 
        \alpha x & \mbox{if}\,\,\,\, x < 0 \\
        x & \,\,\mbox{otherwise}
    \end{cases}
\]
Hence, the output is always small and non-zero when $x<0$. The value of alpha is a predetermined hyperparmater. The effect of the alpha value on the recall is explored in Section \ref{training}. A hyperparameter is a type of parameter whose value is chosen before learning begins. Other parameters are derived during training.

Pooling layers are used to reduce the dimensions of the data by combining the output of a neuron cluster in one layer to a single neuron in the next. The FC layer connects the neurons to all the activations in the previous layer. The final layer is a non-linear classification layer and assigns decimal probabilities to each mutually exclusive class. 

Before running the images through CNN, we scale down the image resolution. The original images had a resolution of 800$\times$400 for the individual Hit and Peak graphs and 800$\times$800 for the images showing both (HitPeak). A lot of computational power would be needed to run these images through the CNN due to the large number of parameters. We chose to scale the HitPeak images to 75x75 resolution and the separate Hit and Peak images to 100x50. Next, the images are converted into a single array, the shape of which depends on the image spatial resolution. Each pixel is one of 256 possible values (0-255) since the images are 8-bit, i.e. 256 = 2$^8$. Most Machine Learning algorithms perform better on small, floating point values instead of large pixel values. Hence, we scale the image pixels to between 0 and 1 by dividing by 255. Each image is labelled as signal or background since we are using supervised learning.

The CNN was created using the {\tt Keras} Sequential model within the {\tt Tensorflow}~\cite{Tensorflow} environment to linearly stack the layers. Each convolution layer contains a 3x3 filter, 16 filters (dimensionality of the output space), with a default stride (how far the filter moves) of 1x1 and no padding. Padding is the process of including an extra layer of zeros around the input image. This is usually done to prevent the convolution output from reducing in size and also to increase the contribution of pixels around the edge of the image. However, since there is no important information at the edges of our images we do not require padding.

An L2 weight regularizer (``kernel\_regularizer") with an alpha value\footnote{Transformation being applied to the coefficients in the weight matrix before being added to the total loss.} of 0.005 is also included in each convolution layer. Regularization is used to prevent overfitting due to intrinsic noise.

Each convolutional layer is followed by a LeakyReLU activation function and a pooling layer. Max pooling, with a pool size of 2x2, is used to reduce the spatial dimensions of the output volume. This means after each pooling layer the output volume is half that of the input volume. 

The classification stage has the most parameters and so requires a ``dropout" regularisation layer to prevent overfitting (with the dropout rate set to 25\%). Dropout is a form of regularisation where each neuron in the CNN has a probability of being deactivated during one iteration. This means a signal cannot pass through a deactivated neuron, causing other neurons to learn features that are not dependent on its surrounds. The probability dropout rate is a hyperparameter and decided before running the model. Dropout only occurs during training; in the testing phase, all neurons are present.

The data is then flattened into a 1-dimensional array to connect the 2-dimensional convolutional and pool layers to the 1-dimension FC layers. The first FC layer (or dense layer) has a unit value of 32. This is the dimension of the output space. The layer also includes both a weight and bias regularizers, both with alpha values 0.001. This layer is again followed by a Leaky ReLU activation function. Another dropout layer is included at this stage, set to 50\%

The output layer contains a single output unit which is used to make predictions. A sigmoid activation function produces a probability output between 0 and 1.

\begin{figure*}[t]
	\begin{center}
		\includegraphics[angle=0,width=1.\linewidth,trim={0mm 0mm 0mm 0mm},clip]{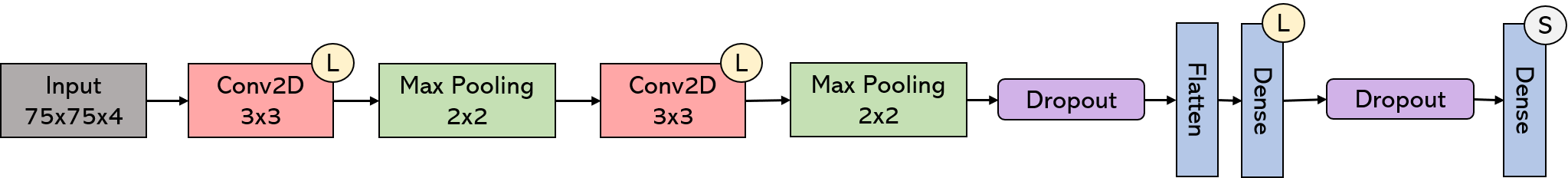}
		\caption{Architecture of CNN with two convolutional layers. Each convolution layer is followed by a LeakyReLU (L) activation function and a Max Pooling layer. The dense layer also contains a LeakyReLU function, while the final output layer contains a Sigmoid (S) activation function. Dropout is used after the final convolution layer (set at 0.25) and before the output layer (set at 0.5)}
		\label{fig:Architecture}
	\end{center}
\end{figure*}

Finally, the model is compiled using an {\tt Adam} optimiser \cite{Kingma:2014vow} to minimise the cost function. {\tt Adam} (derived from `adaptive moment estimation') computes adaptive learning rates for each parameter and has a default learning rate of 0.001. Various learning rates are tested in Section \ref{training}. Binary cross entropy was used for the loss function (since we were completing a binary classification task). Fig. \ref{fig:Architecture} shows the basic architecture of the CNN. 

Finally, the model is fitted with the defined hyperparameters and the accuracy, loss, precision and recall output are recorded. We used recall as the main metric for analysis. Recall (and precision) are preferred over accuracy since they demonstrate the class sensitivity and reliability of the classification of the model. Recall is the ratio of correctly predicted positive (WIMP) observations to all observations in the positive class:
\begin{equation}
    \mbox{Recall} = \frac{TP}{TP + FN}
\end{equation}
\noindent where \textit{TP} is the number of true positives and \textit{FN} is the number of false negatives.

Precision is the ratio of correctly predicted positive observations to the total number of predicted positive observations:
\begin{equation}
    \mbox{Precision} = \frac{TP}{TP + FP}
\end{equation}
\noindent where \textit{FP} is the number of false positives.

\subsection{Training the CNN\label{training}}
We simulated 10,000 events for both WIMP and ER particles, making a total of 20,000 HitPeak training images. The CNN assigns random weights and biases at the beginning of each iteration and so there will always be a slight variation in results after each run. Therefore, the model will give slightly different results in each test, even when using the same parameters. The CNN was run in a GPU {\tt TensorFlow} environment. Since the `HitPeak' images have a different input dimension than the `Hit' and `Peak' images (75x75 instead of 100x50), and the `Hit' and `Peak' images are very different to each other, we need to use a separate CNN model for each image type. We chose to focus on the `HitPeak' images as they contain the most information out of the three image types. Hence, in this section, any mentions of images or dataset refer to the HitPeak data. The results using Hit and Peak images are discussed in Section \ref{Results}.

First we looked at finding the ``optimal" hyperparameters using a grid search. This involves defining a range of values for each hyperparameter in a `grid' which is then used to construct all the possible combinations of the defined hyperparameter values. The CNN then goes through each combination and returns the highest score achieved and the hyperparameter values responsible. 

In our code we used \textit{Stratified k-Fold Cross-Validation}. Cross-validation is used to evaluate machine learning models by splitting the training data into training and validation datasets. The model can then be assessed on how well it will perform on unseen data (such as an unseen test dataset). A k-Fold cross-validator splits the training data into \textit{k} consecutive folds. This allows \textit{k-1} folds to act as the training data with each fold being used once as the validation data. A Stratified cross-validator ensures each fold has the same proportion of observations from each class. In our tests we used 3-fold Stratified Cross-Validation.

In our first test we looked at how the number of convolution layers, epochs and batch size affect the model. 

Having multiple convolutional layers allows a model to extract more complex features. The earlier layers (closer to the input image) learn the lower level features, with the complexity of the features increasing with each layer. To observe how the number of convolutional layers affect our model, CNNs with 1, 2 or 3 layers were used.

As the number of convolutional layers increases, the total number of parameters decreases. This is because there is a pooling layer associated with each convolutional layer which reduces the shape of the output volume. Hence, even though each convolution layer introduces new parameters, the number due to the output volume (when the data is flatten into a 1D array) will be much less than for a CNN with one convolutional layer. This behaviour is only true for architectures which use a flatten layer to generate the input for the final decision layer.

One epoch occurs when the entire dataset in passed through the neural network (via backpropagation) once. The choice of epoch number can lead to overfitting or underfitting of training data depending on whether there are too many or too few epochs. The effect of epoch number is different for different datasets and depends on how diverse the data is (more diverse data requires more epochs so the CNN has time to learn all the data features). We chose to test our CNN using epoch numbers between 20 and 60. In preliminary tests with higher epochs (100 and 150) we found they gave slightly worse results. This could be because our images are relatively simple and so don't have many data features to learn.

The batch size is the number of training data points used in one forward/backward pass. The smaller the batch size, the faster the model will converge to a ``good" solution. This is because the model will start learning before seeing all the data. However, this does not guarantee the model will converge to the best possible result (as it would when using the whole dataset for a binary classification problem, although this is not necessarily true for more than two classes).

\begin{figure}[htp]
    \centering
    \includegraphics[angle=0,width=0.9\linewidth,trim={0mm 0mm 0mm 0mm},clip]{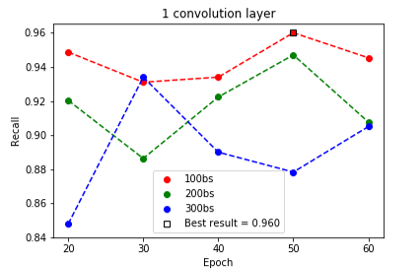}
    \caption{Grid search results for 1 convolution layer; Comparing how batch size (100, 200, 300) and epoch number (20, 30, 40, 50, 60) affect recall for HitPeak training data. The highest recall was achieved using 50 epochs and 100 batch size.}
    \label{fig:1conv}
\end{figure}

\begin{figure}[htp]
    \centering
    \includegraphics[angle=0,width=0.9\linewidth,trim={0mm 0mm 0mm 0mm},clip]{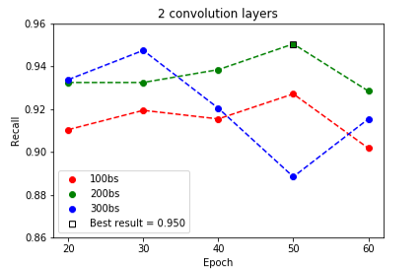}
    \caption{Grid search results for 2 convolution layers; Comparing how batch size (100, 200, 300) and epoch number (20, 30, 40, 50, 60) affect recall for HitPeak training data. The highest recall was achieved using 50 epochs and 200 batch size.}
    \label{fig:2conv}
\end{figure}

\begin{figure}[htp]
    \centering
    \includegraphics[angle=0,width=0.9\linewidth,trim={0mm 0mm 0mm 0mm},clip]{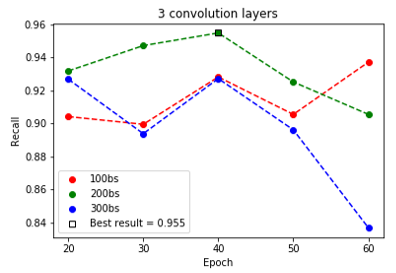}
    \caption{Grid search results for 3 convolution layers; Comparing how batch size (100, 200, 300) and epoch number (20, 30, 40, 50, 60) affect recall for HitPeak training data. The highest recall was achieved using 40 epochs and 200 batch size.}
    \label{fig:3conv}
\end{figure}

Fig. \ref{fig:1conv}-\ref{fig:3conv} show the recall results of the grid search. The hyperparameters tested were number of convolution layers (1, 2 or 3); number of epochs (20, 30, 40, 50 or 60); and batch size (100, 200 or 300). Since we used cross-validation with three folds, each combination was tested three times. Hence the values that are plotted are the averages of those three results. The highest average recall value recorded was 0.960 (3sf) when the CNN had one convolution layer, a batch size of 100 and runs for 50 epochs. Hence, these hyperparameters were used to produce the final CNN model. It is worth noting that all configurations gave a recall between 87\% and 96\%.

\newpage 
There are multiple other different hyperparameters that can be tested and adjusted to improve the model performance, such as the type of optimiser, number of units in the dense layers or number of kernals in the convolution layer. In this experiment we focused on whether the LeakyReLU alpha value and the optimiser learning rate affect the recall ability of the model. In these tests we used the same process as before; a 3-fold stratified cross validation grid search. We used the ``optimal" hyperparameters found before; one convolution layer over 50 epochs with a batch size of 100.

\begin{figure*}
	\begin{center}
	    \includegraphics[scale=0.58]{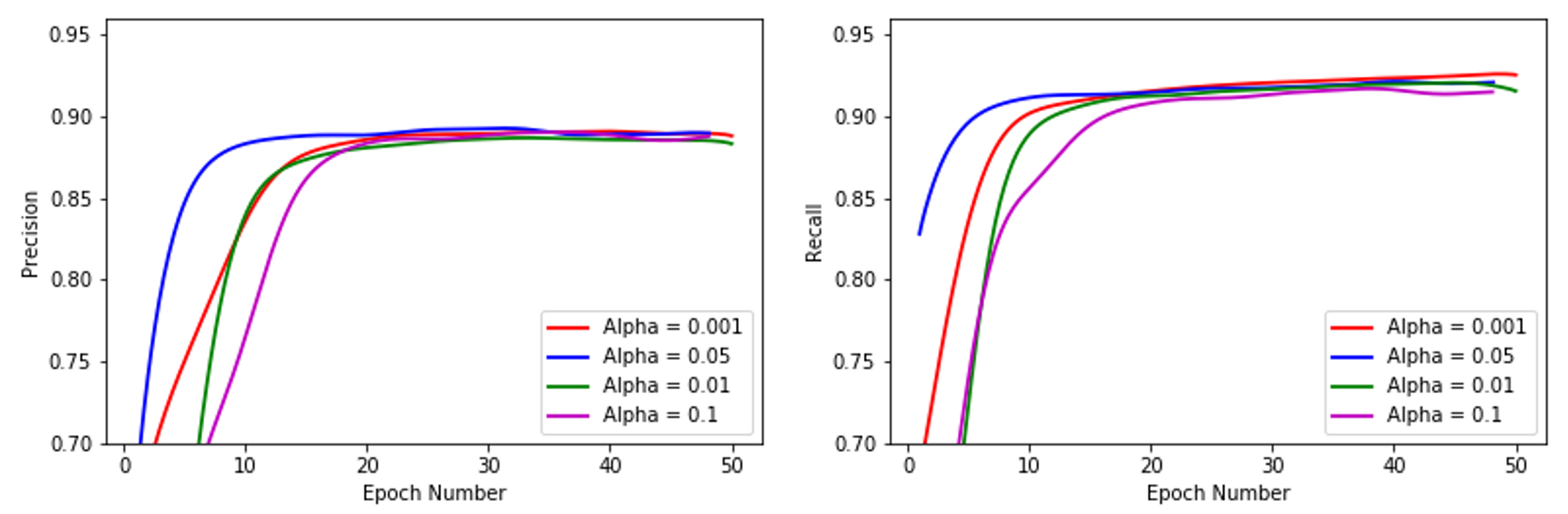}
		\caption{Precision and Recall results for different Leaky ReLU alpha values (0.001, 0.05, 0.01 and 0.1) using HitPeak training data over 50 epochs with a batch size of 100.}
		\label{fig:Alpha}
	\end{center}
\end{figure*}

\begin{figure*}
	\begin{center}
	    \includegraphics[scale=0.58]{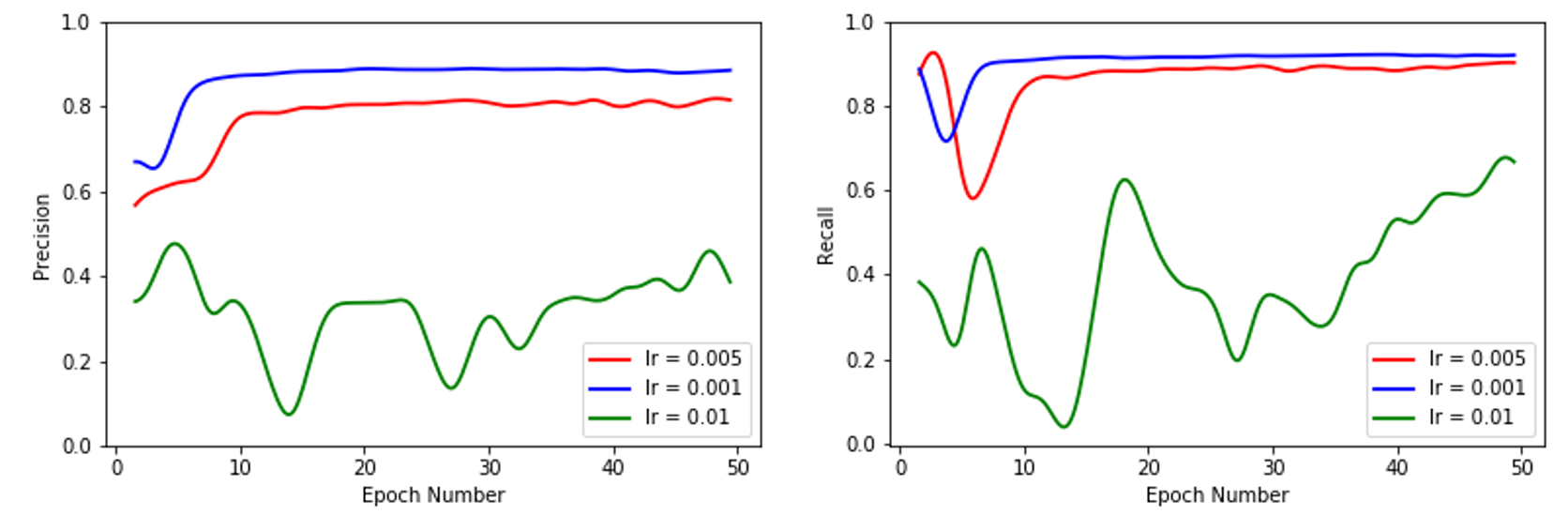}
		\caption{Precision and Recall results for different learning rates (0.005, 0.001, 0.01) using HitPeak training data over 50 epochs with a batch size of 100.}
		\label{fig:lr}
	\end{center}
\end{figure*}

During the previous tests we used an alpha value of 0.05 and learning rate of 0.001. Table \ref{tab:Alpha} shows the recall results when using different alpha values (0.001, 0.05, 0.01 and 0.1), while Fig. \ref{fig:Alpha} presents the precision and recall results. We can see that 0.05 gave the highest final recall score (Table \ref{tab:Alpha}) and reached its `maximum' result is the lowest number of epochs (Fig. \ref{fig:Alpha}). Fig. \ref{fig:lr} shows the accuracy, loss, precision and recall using three different learning rates (0.005, 0.001 and 0.01). Table \ref{tab:lr} gives the final recall scores for each learning rate. We can see that a learning rate of 0.001 gave by far the best results and hence is the best choice for this CNN.

\begin{table}[]
    \centering
    \begin{tabular}{|c|c|}
        \hline
        Alpha Value & Recall \\
        \hline
        0.001 & 0.909 \\
        0.05 & 0.951 \\
        0.01 & 0.907 \\
        0.1 & 0.923 \\
        \hline
    \end{tabular}
    \caption{Recall cross validation values for the four different Leaky ReLU alpha values tested (0.001, 0.05, 0.01, 0.1) using HitPeak training data. An alpha value of 0.05 gave the best result.}
    \label{tab:Alpha}
\end{table}

\begin{table}[]
    \centering
    \begin{tabular}{|c|c|}
        \hline
        Learning Rate & Recall \\
        \hline
        0.005 & 0.933 \\
        0.001 & 0.953 \\
        0.01 & 0.667 \\
        \hline
    \end{tabular}
    \caption{Recall cross validation values for the three different learning rates tested (0.005, 0.001, 0.01) using HitPeak training data. An learning rate of 0.001 gave the best result.}
    \label{tab:lr}
\end{table}

\subsection{Results\label{Results}}
\begin{figure*}[!htbp]
	\begin{center}
	    \includegraphics[scale=0.6]{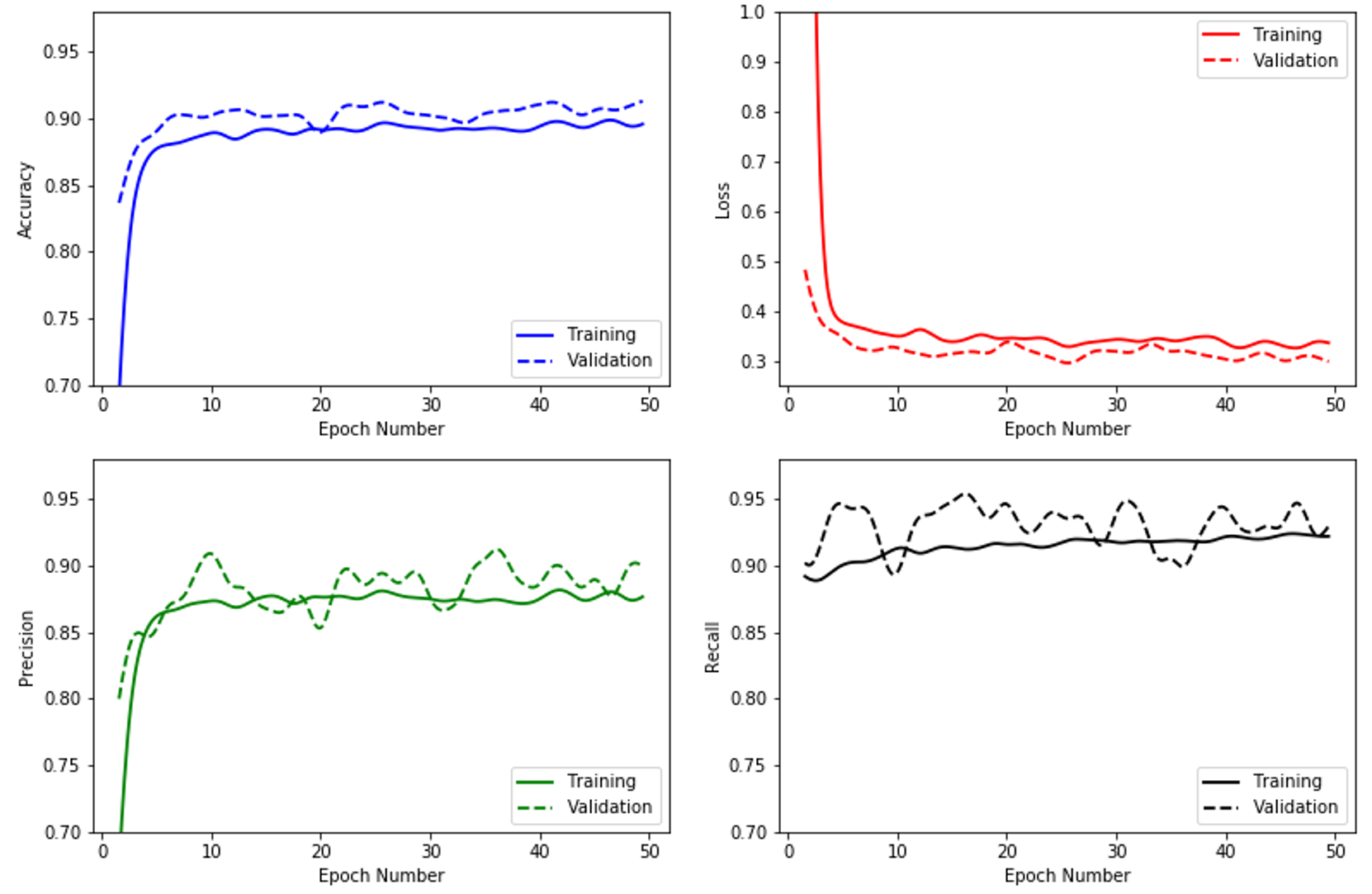}
		\caption{Accuracy, Loss, Precision and Recall results for our final CNN model using the HitPeak training and validation data.}
		\label{fig:Results}
	\end{center}
\end{figure*}

As mentioned at the beginning of the previous section (\ref{training}), each image type (Hit, Peak and HitPeak) requires its own separate CNN model due to the nature of the images. To create the final models, the CNN was set to the ``optimal" hyperparameters found in the previous section. The training data (20,000 images) was split into a training set and a validation set at a 70:30 ratio (14,000 training and 6,000 validation). The CNN was run one more time using these datasets and the final models (including the best weights and biases) for each image type were saved. Fig. \ref{fig:Results} shows the accuracy, loss, precision and recall results for the training and validation data for the final run using the HitPeak images. 

Next the three final models were used on a test data set which included 10,000 newly simulated images (5,000 of each WIMP and ER). The final test doesn't involve any cross-validation or dropout regularisation as the model isn't learning from the new data.

\begin{figure}[!htbp]
	\begin{center}
	    \includegraphics[scale=0.6]{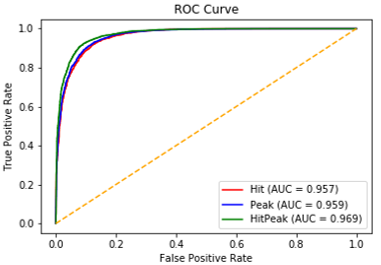}
		\caption{ROC Curve and AUC scores for the three image types (Hit, Peak and HitPeak) using the test data on the final model.}
		\label{fig:Roc}
	\end{center}
\end{figure}

\begin{figure}[!htbp]
	\begin{center}
	    \includegraphics[scale=0.6]{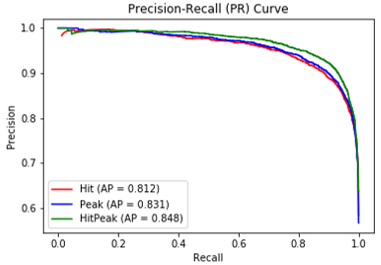}
		\caption{Precision-Recall (PR) Curve and average precision (AP) scores for the three image types (Hit, Peak and HitPeak) using the test data on the final model.}
		\label{fig:PR}
	\end{center}
\end{figure}

\begin{figure}[!htbp]
	\begin{center}
	    \includegraphics[scale=0.8]{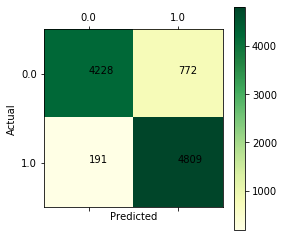}
		\caption{Confusion matrix for the HitPeak test images using the final model. Showing a final recall of 96.18\% with a precision of 86.15\% and an accuracy of 90.37\%.}
		\label{fig:Confusion}
	\end{center}
\end{figure}

\begin{table}[]
    \centering
    \begin{tabular}{|c|c|c|c|}
        \hline
        Metric & Hit (3sf) & Peak (3sf) &  HitPeak (3sf)\\
        \hline
        Accuracy & 0.872 & 0.889 & 0.904 \\
        Recall & 0.934 & 0.937 & 0.962 \\
        Average Precision & 0.812 & 0.831 & 0.848 \\
        \hline
    \end{tabular}
    \caption{Accuracy, Recall and Average Precision results when using the final model on the test dataset for each image type.}
    \label{tab:Final results}
\end{table}

Fig. \ref{fig:Roc} presents the resulting ROC Curve and AUC Scores for each image type, while Fig. \ref{fig:PR} shows their precision-recall curves and average precision. Fig. \ref{fig:Confusion} shows the confusion matrix for the HitPeak images. Table \ref{tab:Final results} gives the final accuracy, recall and average precision of the test data for the three image types. 

From the ROC and PR curves we can see the HitPeak images gave the best results. This could be because the CNNs hyperparameters were chosen based on the training results which used the HitPeak images.

If WIMP acceptance rate is set to 50\% the CNN rejects 99.3\% of all ER events. Hence, the background rate is 3.5 higher than using the standard method mentioned in Section~\ref{NoEvents}.

\section{Conclusions and Outlook\label{conclusions}}
In this work, we have introduced a new method of differentiating between proposed Dark Matter (WIMP) events and the background events (electronic recoils) in the XENON1T experiment using the graphical output of PaX. The final CNN model correctly identified a proposed WIMP event 90.4\% of the time, with a recall of 96.2\% and a precision of 86.2\%. 

Using the grid search method with stratified k-fold cross-validation we found a CNN with one convolution layer run over 50 epochs with a batch size of 100 gave the highest recall of WIMP training HitPeak images (96.0\%). We also showed our choice of learning rate (0.001) and leaky ReLU alpha value (0.05) gave the best results.
This set-up could be used for the real Dark Matter and background models created by the XENON group to simulate more realistic events. The CNN can then be adapted and improved using this data, with the aim of running CNN using real data generated during the XENON1T experiment. Further developments will increase the ER rejection rate of 99.3\%, and possibly improve on the current value of 99.8\% used by XENON1T. If this stage is successful then the method could be used in the next XENON collaboration experiment; the XENONnT (an upgrade of the XENON1T).

Furthermore, it may be possible to use the hitpattern images and a CNN in a regression based analysis to reconstruct the  position of the interaction, instead of the neural network (\cite{Aprile:2019bbb}) currently used in XENON1T.

Experiments aiming to detect Dark Matter via direct detection focus on the particular regions of the detector to improve the signal and background ratio. This method does not require any reduction in the data sets, but just differentiates the events based on the PMTs signals. This idea could also be used for other TPCs based detectors. Finally, this work could be further extended by the ML based models to identify the rare WIMP events among the many different type of background events, not just the dominant ER background.

\section{acknowledgments}
C.K.K. wishes to acknowledge support from the Royal Society-SERB Newton International Fellowship (NF171488). 
The work of V.S. is supported by the Science Technology and Facilities Council (STFC) under grant number $\text{ST/P000819/1}$. We would like to thank Michael Soughton and Adam Matthews for many fruitful discussions. We would also like to acknowledge Prof. Christopher Tunnell and Dr Jelle Aalbers for their help in explaining how we could use PaX, and the other open-source software they created, to carry out this work.\\

\appendix
\newcommand{\hbAppendixPrefix}{A}
\renewcommand{\thefigure}{\hbAppendixPrefix\arabic{figure}}
\setcounter{figure}{0}

\setcounter{table}{0}
\renewcommand{\thetable}{B\arabic{table}}
\section{Analysis set-up\label{setup}}
This section follows the Linux installation process suggested by the PAX github page \cite{PAX}.

\begin{enumerate}
\item Install Python 3 and Anaconda libraries

\tt{wget http://repo.continuum.io/archive/...
...Anaconda3-2.4.0-Linux-x86\_64.sh \\
bash Anaconda3-2.4.0-Linux-x86\_64.sh}

\item Set up Anaconda libraries

\texttt{\noindent export PATH=$\sim$/anaconda3/bin:\$PATH \\
\noindent conda config --add channels http://conda.anaconda.org/NLeSC}

\item Add additional python packages

\texttt{conda install conda \\
conda create -n pax python=3.4 root=6 numpy scipy=0.18.1 pyqt=4.11 matplotlib pandas cython h5py numba pip python-snappy pytables scikit-learn rootpy pymongo psutil jupyter dask root\_pandas jpeg=8d isl=0.12.2 gmp=5.1.2 glibc=2.12.2 graphviz=2.38.0=4 gsl=1.16 linux-headers=2.6.32 mpc=1.0.1 mpfr=3.1.2 pcre=8.37 python-snappy=0.5 pyopenssl=0.15.1} 
	
\item Activate the environment

\texttt{source activate pax }

\item Installing Pax

\texttt{git clone https://github.com/XENON1T/pax.git \\
source activate pax\\
cd pax\\
python setup.py develop}

\end{enumerate}

\section{Simulate Data - Package Installation}

\underline{Laidbax} - \cite{Laidbax}

\texttt{\indent git clone https://github.com/XENON1T/laidbax\\
\indent \indent cd laidbax\\
\indent \indent python setup.py develop\\
\indent \indent cd ..}

\underline{Blueice} - \cite{Blueice}

\texttt{\indent git clone https://github.com/JelleAalbers/blueice\\
\indent \indent cd blueice\\
\indent \indent python setup.py develop\\
\indent \indent cd ..}

\underline{Wimprates} - \cite{wimprate}
\texttt{\indent pip install wimprates} 

\underline{Multihist} - \cite{multihist}
\texttt{\indent pip install multihist }

\newpage

\section{Simulate Data - Running}

The code to simulate and plot the data can be found in \cite{lucygithub}.
\begin{itemize}
	\item Run ``\texttt{python Simulate.py}" (This python file is based on the laidbax tutorial notebook \cite{Laidbax}). This will save two csv files called `ERSIM.csv' and `WIMPSIM.csv' in the data folder ($\sim$/pax/pax/data)	
	\item To generate the hitpattern and peak images run:
	``\texttt{paxer --config XENON1T Simulation --input $\sim$/pax/pax/data/WIMPSIM.csv --plot}"
    \item Replace `WIMPSIM' with `ERSIM' for ER plots.
\end{itemize}



\section{Running the CNN}

To run the CNN create a conda environment containing TensorFlow:\\
\indent \indent \texttt{conda create -n tensorflow\_env tensorflow}\\
\indent \indent \indent	\texttt{conda activate tensorflow\_env} \\

Then install ``Keras" \cite{Keras}, ``Matplotlib" \cite{Matplotlib}, ``Scikit-learn" \cite{Sklearn}, ``Pandas" \cite{Pandas} and ``Imageio" \cite{Imageio}: \\
\indent \indent \texttt{conda install -c conda-forge keras}\\
\indent \indent \texttt{conda install -c conda-forge matplotlib}\\
\indent \indent \texttt{conda install scikit-learn}\\
\indent \indent \texttt{conda install -c anaconda pandas}\\
\indent \indent \texttt{conda install -c menpo imageio}


\begin{thebibliography}{99}

\bibitem{Agnes:2018fwg}
  P.~Agnes {\it et al.} [DarkSide Collaboration],
  Phys.\ Rev.\ D {\bf 98} (2018) no.10,  102006,
  [arXiv:1802.07198 [astro-ph.CO]].

\bibitem{Akerib:2016vxi}
  D.~S.~Akerib {\it et al.} [LUX Collaboration],
  Phys.\ Rev.\ Lett.\  {\bf 118} (2017) no.2,  021303,
  [arXiv:1608.07648 [astro-ph.CO]].

\bibitem{Cui:2017nnn}
  X.~Cui {\it et al.} [PandaX-II Collaboration],
  Phys.\ Rev.\ Lett.\  {\bf 119} (2017) no.18,  181302,
  [arXiv:1708.06917 [astro-ph.CO]].
  
\bibitem{Aprile:2017aty}
  E.~Aprile {\it et al.} [XENON Collaboration],
  Eur.\ Phys.\ J.\ C {\bf 77} (2017) no.12,  881,
  [arXiv:1708.07051 [astro-ph.IM]].
  
\bibitem{Carleo:2019ptp}
  G.~Carleo, I.~Cirac, K.~Cranmer, L.~Daudet, M.~Schuld, N.~Tishby, L.~Vogt-Maranto and L.~Zdeborová,
  arXiv:1903.10563 [physics.comp-ph].


\bibitem{Zhang:2019ryt}
  X.~Zhang, Y.~Wang, W.~Zhang, Y.~Sun, S.~He, G.~Contardo, F.~Villaescusa-Navarro and S.~Ho,
  arXiv:1902.05965 [astro-ph.CO].
 
\bibitem{Fluri:2019qtp}
  J.~Fluri, T.~Kacprzak, A.~Lucchi, A.~Refregier, A.~Amara, T.~Hofmann and A.~Schneider,
  Phys.\ Rev.\ D {\bf 100} (2019) no.6,  063514,
  [arXiv:1906.03156 [astro-ph.CO]]. 
  
\bibitem{Khosa:2019kxd}
  C.~K.~Khosa, V.~Sanz and M.~Soughton,
  arXiv:1910.06058 [hep-ph].
  
\bibitem{Brehmer:2019jyt}
  J.~Brehmer, S.~Mishra-Sharma, J.~Hermans, G.~Louppe and K.~Cranmer,
  arXiv:1909.02005 [astro-ph.CO].


\bibitem{Alexander:2019puy}
  S.~Alexander, S.~Gleyzer, E.~McDonough, M.~W.~Toomey and E.~Usai,
  arXiv:1909.07346 [astro-ph.CO].

\bibitem{DiazRivero:2019hxf}
  A.~Diaz Rivero and C.~Dvorkin,
  arXiv:1910.00015 [astro-ph.CO].


\bibitem{cosmoDM}
  C.~Escamilla-Rivera, M.~A.~C.~Quintero and S.~Capozziello,
  arXiv:1910.02788 [astro-ph.CO].  

\bibitem{Aprile:2010bt}
  E.~Aprile {\it et al.} [XENON Collaboration],
  Astropart.\ Phys.\  {\bf 34} (2011) 679
  doi:10.1016/j.astropartphys.2011.01.006
  [arXiv:1001.2834 [astro-ph.IM]].

\bibitem{Aprile:2011dd}
  E.~Aprile {\it et al.} [XENON100 Collaboration],
  Astropart.\ Phys.\  {\bf 35} (2012) 573,
  [arXiv:1107.2155 [astro-ph.IM]].

\bibitem{Aprile:2017iyp}
  E.~Aprile {\it et al.} [XENON Collaboration],
  Phys.\ Rev.\ Lett.\  {\bf 119} (2017) no.18,  181301,
  [arXiv:1705.06655 [astro-ph.CO]].

\bibitem{Pelssers}
B. Pelssers, \textit{Position reconstruction and data quality in Xenon}, Master’s thesis, Universiteit Utrecht (2015).

\bibitem{WIMPbanchmark} {Bagnaschi E et al 2019 arXiv:1905.00892 [hep-ph]}

\bibitem{Laidbax}
Laidbax Github repository, XENON Collaboration https://github.com/XENON1T/laidbax

\bibitem{Blueice}
J. Aalbers, Moraa K Pelssers B 2019 JelleAalbers/blueice: v1.0.0-beta.2  (Version v1.0.0-beta.2) Zendo http://doi.org/10.5281/zendo.3234038

\bibitem{Aalbers:2018mfc}
  J.~Aalbers, \textit{Dark matter search with XENON1T}, PhD thesis, University of Amsterdam (2018).

\bibitem{PAX}
PaX Github repository, XENON Collaboration, https://github.com/XENON1T/pax.

\bibitem{Simola:2018ntn}
  U.~Simola, B.~Pelssers, D.~Barge, J.~Conrad and J.~Corander,
  JINST {\bf 14} (2019) no.03,  P03004,
  [arXiv:1810.09930 [astro-ph.IM]].  
 
\bibitem{wimprate}
J. Aalbers, B. Pelssers and K. Morå, \textit{wimprates: v0.3.0} (2019). http://doi.org/10.5281/zenodo.3345959.

\bibitem{Lenardo:2014cva}
  B.~Lenardo, K.~Kazkaz, A.~Manalaysay, J.~Mock, M.~Szydagis and M.~Tripathi,
  IEEE Trans.\ Nucl.\ Sci.\  {\bf 62} (2015) no.6,  3387,
  [arXiv:1412.4417 [astro-ph.IM]].

\bibitem{Breur}
P. A. Breur, \textit{Backgrounds in XENON1T}, PhD thesis, University of Amsterdam (2019).

\bibitem{Aprile:2019dme}
  E.~Aprile {\it et al.} [XENON Collaboration],
  Phys.\ Rev.\ D {\bf 99} (2019) no.11,  112009
  doi:10.1103/PhysRevD.99.112009
  [arXiv:1902.11297 [physics.ins-det]].


\bibitem{Aprile:2018dbl}
  E.~Aprile {\it et al.} [XENON Collaboration],
  Phys.\ Rev.\ Lett.\  {\bf 121} (2018) no.11,  111302
  doi:10.1103/PhysRevLett.121.111302
  [arXiv:1805.12562 [astro-ph.CO]].
  
  
\bibitem{lucygithub}
L. Mars, Github repository for this paper, https://github.com/LucyMars/SearchForDarkMatter (2020).

\bibitem{Tensorflow}
M. Abadi, et al., ``TensorFlow: Large-scale Machine Learning on heterogeneous distributed systems" arXiv:1603.04467 (2016).


\bibitem{Kingma:2014vow}
  D.~P.~Kingma and J.~Ba,
  arXiv:1412.6980 [cs.LG].

\bibitem{Aprile:2019bbb}
  E.~Aprile {\it et al.} [XENON Collaboration],
  Phys.\ Rev.\ D {\bf 100} (2019) no.5,  052014
  doi:10.1103/PhysRevD.100.052014
  [arXiv:1906.04717 [physics.ins-det]].

  
\bibitem{multihist}
J. Aalbers, Multihist github repository https://github.com/JelleAalbers/multihist

\bibitem{Keras}
F. Chollet, Keras GitHub repository, https://github.com/fchollet/keras (2015).

\bibitem{Matplotlib}
J. Hunter, ``Matplotlib: A 2D Graphics Environment", \textit{Computing in Science and Engineering}, \textbf{9}, 90-95 (2007).

\bibitem{Sklearn}
F. Pedregosa, et al., ``Scikit-learn: Machine learning in Python", \textit{Journal of Machine Learning Research}, \textbf{12}, pp. 2825-2830 (2011).

\bibitem{Pandas}
W. McKinney, ``Data Structures for Statistical Computing in Python", \textit{Proceedings of the 9th Python in Science Conference}, 51-56 (2010).

\bibitem{Imageio}
A. Klein, et al., imageio/imageio: V2.6.0 (Version v2.6.0). Zenodo. http://doi.org/10.5281/zenodo.3475011 (2019).



\end{thebibliography}
\end{document}